\documentclass[english,prb,twocolumn]{revtex4-2}
\usepackage[T1]{fontenc}
\usepackage[latin9]{inputenc}
\usepackage{amsmath}
\usepackage{amssymb}
\usepackage{graphicx}
\usepackage{babel}
\begin{document}
\title{Controlled phase gate in exchange coupled quantum dots affected by
quasistatic charge noise}
\author{Yinan Fang$^{1,2}$}
\email{ynfang@ynu.edu.cn}

\affiliation{$^{1}$School of Physics and Astronomy and Yunnan Key Laboratory for
Quantum Information, Yunnan University, Kunming 650500, China~\\
$^{2}$Beijing Computational Science Research Center, Beijing 100193,
People's Republic of China}
\begin{abstract}
Charge noise has been one of the main issues in realizing high fidelity
two-qubit quantum gates in semiconductor based qubits. Here, we study
the influence of quasistatic noise in quantum dot detuning on the
controlled phase gate for spin qubits that defined on a double quantum
dot. Analytical expressions for the noise averaged Hamiltonian, exchange
interaction, as well as the gate fidelity are derived for weak noise
covering experimental relevant regime. We also perform interleaved
two-qubit randomized benchmarking analysis for the controlled phase
gate and show that an exponential decay of the sequential fidelity
is still valid for the weak noise.
\end{abstract}
\maketitle

\section{Introduction}

Implementing high fidelity gates driven by the pursuit for a large
scale quantum computational device had been witnessed with remarkable
progresses during the past decades. In spin qubits based on the semiconductor
quantum dot (QD) \citep{1998_PRA_Loss,2007_RMP_Hanson}, single as
well as two qubits gates fidelity exceeding or close to the threshold
of quantum error correction on the surface code \citep{1998_arXiv_Bravyi,2002_JMathPhys_Dennis,2012_PRA_Fowler}
had been routinely achieved \citep{2014_NatNano_Veldhorst,2018_NatNano_Yoneda,2019_Nature_Huang}.
Meanwhile, architectures for a scalable design were proposed for spin
qubits that based on the phosphorus donor \citep{2015_SciAdv_HillCharles}
and QDs \citep{2017_NatCommun_Veldhorst,2018_SciAdv_Li,2021_SciRep_Tadokoro}.
Recently, coherent control and spin readout were demonstrated for
a three-by-three two-dimensional QD array in GaAs/AlGaAs \citep{2021_NatNano_Mortemousque},
and devices operated at a higher temperature enabling larger cooling
power were also realized in isotopically purified silicon \citep{2022_CommunMater_Petit}. 

In contrast to single qubit gates usually implemented by applying
with an oscillating magnetic field, a natural way to implement the
two-qubit operation in gate defined in a double quantum dot (DQD)
is to utilize the Heisenberg exchange interaction $J$ \citep{1998_PRA_Loss,1999_PRB_Burkard,2011_PRB_Meunier}.
Usually, electric noise affects $J$ in DQD more strongly than those
parameters that involved in single qubit gates, as it depends significantly
on the confining potential and the overlap of the electron wave functions
\citep{1999_PRB_Burkard,2006_PRL_Hu}. Furthermore, the suppression
of magnetic noise using techniques such as dynamical decoupling and
isotopically purification makes the charge noise the current bottleneck
in order to reach threshold fidelity for two-qubit gates error correction
\citep{2018_NatNano_Yoneda}. Also the readout fidelity was shown
to be limited by electric noise \citep{2019_NatNano_Urdampilleta}.
Therefore, a better understanding on the decoherence effect of charge
noise accounting for the state-of-art progress in QD based spin qubits
would be beneficial in further boosting the two-qubit gate fidelities
towards the ultimate goal of quantum error correction.

The source of charge noise could either be mobile or fixed \citep{2020_PRAppl_Chanrion}.
One typical example of mobile charge noise is due to fluctuating two-level
systems near the QDs, while the fixed charge noise is due to static
disorders. In this regards, a usually employed approach to study the
detrimental effect of charge noise in QD is to assume that the noise
affects the system only quasistatically, where the typical frequency
of the noise is smaller than $1/T_{2}^{*}$ \citep{2013_PRL_Dial}.
Under such assumption, $J$ is taken as a fixed quantity during each
implementation of a quantum gate but can vary randomly between successive
execute of the gate, then physical quantities are calculated by average
over the distribution of $J$. The quasistatic model of charge noise
had been applied in the calculation of the electron spin inhomogeneous
dephasing time $T_{2}^{*}$ \citep{2017_PRB_Throckmorton,2021_PRB_Buterakos,2022_npjQuantumInfor_Keith,2022_PRB_Throckmorton},
and it reveals effects such as the transition from power-law to exponential
decay in free induced decay signals \citep{2013_PRL_Dial,2022_npjQuantumInfor_Keith}.

In this work, we investigate theoretically the influence of charge
noise on the controlled phase (CPHASE) gate in a DQD. Instead of imposing
phenomenological noise distributions on $J$, we rely on a microscopic
model which accounts for the dependence of $J$ on these parameters
that are directly connected to charge fluctuations \citep{2022_npjQuantumInfor_Keith}.
Perturbative results on the experimental relevant quantities such
as the spin-up fraction and the gate fidelity are derived in the limit
of weak noise, which covers the regime of experimental parameters
\citep{2014_NatNano_Veldhorst,2015_Nature_Veldhorst,2019_Nature_Huang}.
We also quantify the effect of charge noise by two-qubit randomized
benchmarking (RB), and the results suggest that the noise averaged
sequential fidelity still decays according to an exponential-law at
the realistic noise strength. 

This paper is organized as follows. In Sec. II, we introduce the model
for the DQD system, together with the quasistatic description for
the charge noise. Meanwhile, the average effect on the model Hamiltonian
is also derived. In Sec. III, the implementation of the CPHASE gate
was discussed, with the derivation of perturbative results for the
spin-up fraction and the gate fidelity. We supplied those results
in Sec. IV by the two-qubit interleaved RB simulation on the CPHASE
gate. Finally, we summarize and draw the conclusion in Sec. V.

\section{The model\label{Sec_II}}

\begin{figure}
\begin{centering}
\includegraphics[width=8.6cm]{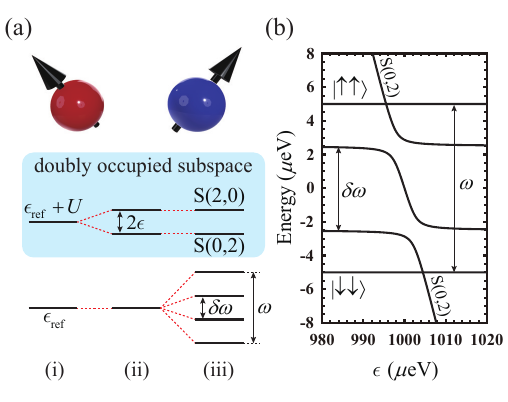}
\par\end{centering}
\caption{(a) Schematic plot of a double quantum dot (DQD) and its energy levels
in the two electrons subspace. The red and blue dots are subjected
to different magnetic fields along the $z$-direction. In the bottom
panel, from left to right the accumulated effects on the level splitting
are: (i) Coulomb repulsion $U$; (ii) Voltage detuning $\epsilon$;
(iii) Zeeman splitting $\omega$ and $\delta\omega$ due to the inhomogeneous
magnetic field. $\epsilon_{\mathrm{ref}}$ is a referential energy
point. (b) Energy levels as a function of the voltage detuning $\epsilon$
near the $\mathrm{S}(0,2)$-$(1,1)$ anti-crossing. Parameters are
chosen for a clear demonstration of the energy levels near the anti-crossing:
$\epsilon_{\mathrm{ref}}=0$, $t=1\mu\mathrm{eV}$, $\omega=10\mu\mathrm{eV}$,
$\delta\omega=5\mu\mathrm{eV}$, and $U=1\mathrm{meV}$.\label{FIG_schematics}}
\end{figure}

Our model consists of a double quantum dot (DQD) subjected to an inhomogeneous
magnetic in the $z$-direction, see FIG. \ref{FIG_schematics} (a).
We describe the model with the Hubbard Hamiltonian \citep{2011_PRB_Meunier},
\begin{align}
\hat{H}= & \sum_{\sigma\in\{\uparrow,\downarrow\}}\left[\sum_{j=1,2}\epsilon_{j,\sigma}\hat{d}_{j,\sigma}^{\dagger}\hat{d}_{j,\sigma}+\left(te^{i\phi}\hat{d}_{1,\sigma}^{\dagger}\hat{d}_{2,\sigma}+\mathrm{h.c.}\right)\right]\nonumber \\
 & +U\sum_{j}\hat{n}_{j,\uparrow}\hat{n}_{j,\downarrow},\label{Hamiltonian}
\end{align}
where $\hat{d}_{j,\sigma}$ annihilates an electron of spin $\sigma$
on dot $j$, and $\hat{n}_{j,\sigma}\equiv\hat{d}_{j,\sigma}^{\dagger}\hat{d}_{j,\sigma}$
is the corresponding 
electron number operator. The first line of Eq. (\ref{Hamiltonian})
describes the combined effects of voltage detuning, Zeeman splitting
due to magnetic field, as well as the tunneling coupling between the
two dots. The second line of Eq. (\ref{Hamiltonian}) describes the
on-site Coulomb repulsion.

In order to better distinguish the detuning and magnetic fields effects
that both contained in $\epsilon_{j,\sigma}$, one could introduce
another set of parameters (setting $\hbar\equiv1$):
\begin{equation}
\epsilon_{j}=\frac{1}{2}\left(\epsilon_{j,\uparrow}+\epsilon_{j,\downarrow}\right),\text{ }\omega_{j}=\epsilon_{j,\uparrow}-\epsilon_{j,\downarrow},
\end{equation}
With $\epsilon_{j}$ and $\omega_{j}$, the detuning of the first
dot with respect to the second dot is given by
\begin{equation}
\epsilon=\epsilon_{1}-\epsilon_{2},
\end{equation}
while the inhomogenous magnetic field leads to the following Zeeman
energy difference between the two dots
\begin{equation}
\delta\omega=\omega_{1}-\omega_{2}.
\end{equation}
The required local magnetic field difference for $\delta\omega$ could
be generated by several methods, by micromagnet \citep{2006_PRL_Tokura,2014_PRB_Chesi,2018_NatNano_Yoneda}
or dynamical nuclear polarization \citep{2014_PRB_Thalineau}, in
silicon QDs by the Stark effect \citep{2014_NatNano_Veldhorst,2015_Nature_Veldhorst},
or in cross-bar architecture by external current \citep{2018_SciAdv_Li}.
For the convenience of analysis we also introduce the total Zeeman
splitting $\omega\equiv\omega_{1}+\omega_{2}$, as well as an energy
referential point $\epsilon_{\mathrm{ref}}=\epsilon_{1}+\epsilon_{2}$. 

The physical meaning of the above parameters can be clearly seen from
the energy spectrum of $\hat{H}$, which is depicted in FIG. \ref{FIG_schematics}
under the two-electron subspace. In the absence of tunneling coupling
$t=0$, the quantities $\omega$ and $\delta\omega$ give the respective
energy spacing between the spin parallel ($|\uparrow,\uparrow\rangle$
and $|\downarrow,\downarrow\rangle$) and spin anti-parallel ($|\uparrow,\downarrow\rangle$
and $|\downarrow,\uparrow\rangle$) subspaces. For realistic consideration,
the reported value of $\omega$ in the DQD based on the isotopically
purified silicon is about $160\mu\mathrm{eV}$ \citep{2015_Nature_Veldhorst,2019_Nature_Huang},
which corresponds to a magnetic field of $1.4\mathrm{T}$. Electrically
tuning of the $g$-factor provides a Zeeman field difference $\delta\omega=0.16\mu\mathrm{eV}$
($\delta\omega/h=40\mathrm{MHz}$) \citep{2014_NatNano_Veldhorst,2019_Nature_Huang}.
Finally, the tunneling coupling about $4\mu\mathrm{eV}$ was obtained
from fitting the ESR spectrum to the Hubbard model, while the Coulomb
repulsion $U$ was estimated to be $220\mathrm{meV}$ in Ref. \citep{2018_PRB_Gungordu}.

The separation of energy 
\begin{figure}
\begin{centering}
\includegraphics[width=8.6cm]{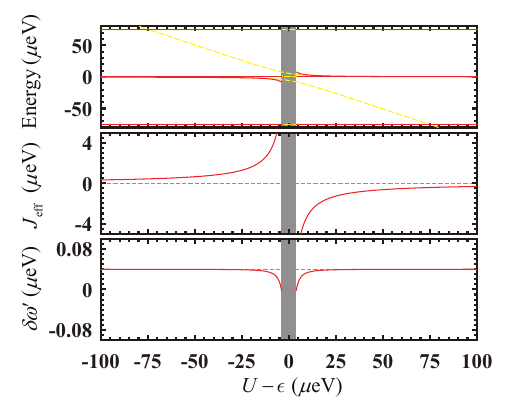}
\par\end{centering}
\caption{Energy spectrum and parameters from the effective Hamiltonian $\hat{H}_{\mathrm{eff}}$
in Eq. (\ref{HamEff}) near the $\mathrm{S}(0,2)$-$(1,1)$ anti-crossing.
The gray shaded region is defined by $|U-\epsilon|\protect\leq|t-\delta\omega/2|$,
inside this region the effective model become invalid. The red curves
are solved from Eqs. (\ref{HamEff},\ref{Jeff},\ref{dwp}) while
the yellow dashed curves in the top panel are energies obtained from
the full Hamiltonian Eq. (\ref{Hamiltonian}). Parameters used in
the calculations: $t=4\mu\mathrm{eV}$, $\epsilon_{\mathrm{ref}}=0$,
$\omega=150\mu\mathrm{eV}$, $\delta\omega=0.04\mu\mathrm{eV}$, and
$U=220\mathrm{meV}$. \label{FIG_EffectiveParameters}}

\end{figure}
scales $U\gg|\delta\omega|,|\omega|$ allows one to derive an effective
description of the DQD. In particular, except when the detuning $\epsilon$
was tuned close to $\epsilon\simeq\pm U$, states with a doubly occupied
dot, i.e., $|\mathrm{S}(2,0)\rangle\equiv\hat{d}_{1,\uparrow}^{\dagger}\hat{d}_{1,\downarrow}^{\dagger}|0\rangle$
or $|\mathrm{S}(0,2)\rangle\equiv\hat{d}_{2,\uparrow}^{\dagger}\hat{d}_{2,\downarrow}^{\dagger}|0\rangle$,
would always energetically be separated from the other four $(1,1)$
states $|\sigma,\sigma'\rangle$ ($\sigma,\sigma'\in\{\uparrow,\downarrow\}$),
thus the two doubly occupied states could be eliminated by the Schrieffer-Wolf
transformation \citep{1964_BOOK_Schrieffer,1966_PR_Schrieffer}, leading
to the following effective Hamiltonian
\begin{align}
\hat{H}_{\mathrm{eff}}= & \epsilon_{\mathrm{ref}}-\frac{J_{\mathrm{eff}}}{4}+J_{\mathrm{eff}}\hat{\mathbf{S}}_{1}\cdot\hat{\mathbf{S}}_{2}+\frac{1}{2}[\omega(\hat{S}_{1}^{z}+\hat{S}_{2}^{z})\nonumber \\
 & +\delta\omega'(\hat{S}_{1}^{z}-\hat{S}_{2}^{z})].\label{HamEff}
\end{align}
Here, the exchange interaction $J_{\mathrm{eff}}$ depends on the
tunneling coupling as well as the detuning as follows \citep{1999_PRB_Burkard,2011_PRB_Meunier}
\begin{equation}
J_{\mathrm{eff}}=t^{2}\left(\frac{1}{\Omega_{\uparrow,\downarrow}}+\frac{1}{\Omega_{\downarrow,\uparrow}}\right),\label{Jeff}
\end{equation}
 where
\begin{equation}
\frac{1}{\Omega_{\sigma,\sigma'}}=\frac{1}{E_{\mathrm{S}(0,2)}-E_{\sigma,\sigma'}}+\frac{1}{E_{\mathrm{S}(2,0)}-E_{\sigma,\sigma'}},
\end{equation}
and $E_{j}$ denotes the energy of the DQD eigenstate $|j\rangle$
in the absence of the tunneling coupling, i.e.,
\begin{equation}
E_{\mathrm{S}(2,0)}=2\epsilon_{1}+U,\text{ }E_{\mathrm{S}(0,2)}=2\epsilon_{2}+U,
\end{equation}
and
\begin{equation}
E_{\uparrow,\downarrow}=\epsilon_{\mathrm{ref}}+\delta\omega/2,\text{ }E_{\downarrow,\uparrow}=\epsilon_{\mathrm{ref}}-\delta\omega/2.
\end{equation}
Because of the tunneling coupling, the Zeeman energy difference $\delta\omega$
is modified to $\delta\omega'$, which is defined as follows
\begin{equation}
\delta\omega'=\delta\omega-t^{2}\left(\Omega_{\uparrow,\downarrow}^{-1}-\Omega_{\downarrow,\uparrow}^{-1}\right).\label{dwp}
\end{equation}
In deriving Eq. (\ref{HamEff}) we have neglected the higher order
terms $\sim o(t^{2}\Omega_{\sigma,\sigma'}^{-2})$, thus the effective
Hamiltonian is valid only when $\epsilon$ is at least $|t-|\delta\omega|/2|$
away from the the $\mathrm{S}(0,2)$-$(1,1)$ anti-crossing point
($\epsilon\equiv U$), this is demonstrated in FIG. \ref{FIG_EffectiveParameters}
by plotting the $J_{\mathrm{eff}}$ and $\delta\omega'$ as a function
of the detuning $\epsilon$ for typical parameters of the DQD.

To account for the charge noise affecting the DQD system, it would
be useful to notice that the time-scale of the noise may span over
a wide range. Therefore, if the noise fluctuates ($\sim\mu\mathrm{s}$)
much slower than the gate operation time ($\sim\mathrm{ns}$), one
may treat the noise as quasistatic \citep{2018_JPC_Ferraro,2021_PRB_Buterakos,2022_PRB_Throckmorton}.
The effect of the noise in this quasistatic limit had been modeled
by assuming a Gaussian distribution for the detuning $\epsilon$ \citep{2017_PRB_Throckmorton,2019_Nature_He,2022_npjQuantumInfor_Keith},
i.e., 
\begin{equation}
p(\epsilon)=\frac{1}{\sqrt{2\pi\sigma_{\epsilon}^{2}}}\exp\left[-\frac{(\epsilon-\bar{\epsilon})^{2}}{2\sigma_{\epsilon}^{2}}\right],\label{Gauss_noise}
\end{equation}
where $\bar{\epsilon}$ represents the 
\begin{figure}
\begin{centering}
\includegraphics[width=8.6cm]{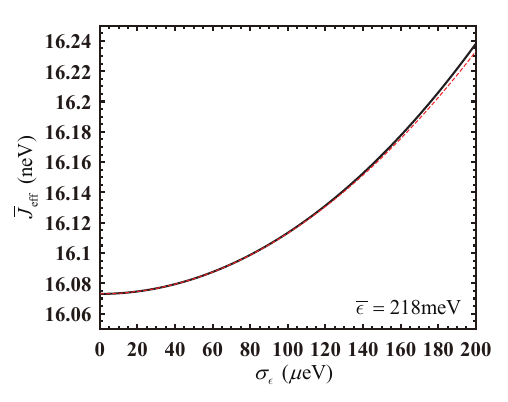}
\par\end{centering}
\caption{Average exchange interaction $\bar{J}_{\mathrm{eff}}$ as a function
of the standard deviation $\sigma_{\epsilon}$ of the distribution
$p(\epsilon)$. Black solid curve is obtained from the numerical integration
while the red dashed curve depicts the approximation Eq. (\ref{Jeffaprx}).
Parameters used in the calculation: $t=4\mu\mathrm{eV}$, $\omega=150\mu\mathrm{eV}$,
$\delta\omega=0.04\mu\mathrm{eV}$, $U=220\mathrm{meV}$, and $\bar{\epsilon}=218\mathrm{meV}$.\label{FIG_JeffApprox}}
\end{figure}
average value of the detuning and the standard deviation $\sigma_{\epsilon}$
quantifies the noise. The value for $\sigma_{\epsilon}$ could be
estimated from the $T_{2}^{*}$ measurements and for the silicon quantum
dot $\sigma_{\epsilon}\simeq28\mu\mathrm{eV}$ \citep{2018_PRB_Gungordu}.

Under the quasistatic limit with the Gaussian distribution model Eq.
(\ref{Gauss_noise}), one can first evaluate the relevant quantities
at a fixed value of $\epsilon$, and the effect of the noise was then
obtained as an average of that quantity with respect to the distribution
$p(\epsilon)$. For example, let us consider here the average of $J_{\mathrm{eff}}$.
If $\bar{\epsilon}$ is far from the $\mathrm{S}(0,2)$-$(1,1)$ anti-crossing
point compared with the width $\sigma_{\epsilon}$ of the distribution
$p(\epsilon)$, one may perform a Taylor expansion on $J_{\mathrm{eff}}(\epsilon)$
around $\bar{\epsilon}$ to the quadratic order $(\epsilon-\bar{\epsilon})^{2}$
and obtain the following approximated expression for the average exchange
interaction
\begin{align}
\bar{J}_{\mathrm{eff}} & =\int d\epsilon J_{\mathrm{eff}}(\epsilon)p(\epsilon)\simeq J_{\mathrm{eff}}(\bar{\epsilon})+\frac{\sigma_{\epsilon}^{2}}{2}J_{\mathrm{eff}}''(\bar{\epsilon}),\label{Jeffaprx}
\end{align}
where $J_{\mathrm{eff}}''(\epsilon)=d^{2}J_{\mathrm{eff}}/d\epsilon^{2}$
is the second order derivative with respect to the detuning. As compared
in FIG. \ref{FIG_JeffApprox}, the approximated $\bar{J}_{\mathrm{eff}}$
given by Eq. (\ref{Jeffaprx}) above agrees quantitatively well with
the exact result provided that $\sigma_{\epsilon}\ll|U-\bar{\epsilon}|$.
For another example, by using the same approximation as in Eq. (\ref{Jeffaprx}),
one can show that the effective Hamiltonian after averaging over $p(\epsilon)$
could be written as follows
\begin{equation}
\bar{H}_{\mathrm{eff}}=\hat{H}_{\mathrm{eff}}+\delta\hat{H}_{\mathrm{c}},
\end{equation}
with {[}neglecting a constant energy shift $-\sigma_{\epsilon}^{2}J_{\mathrm{eff}}''(\bar{\epsilon})/8${]}
\begin{align}
\delta\hat{H}_{\mathrm{c}}= & \lambda_{\uparrow,\downarrow}\left[\hat{\mathbf{S}}_{1}\cdot\hat{\mathbf{S}}_{2}-\left(\hat{S}_{1}^{z}-\hat{S}_{2}^{z}\right)\right]\nonumber \\
 & +\lambda_{\downarrow,\uparrow}\left[\hat{\mathbf{S}}_{1}\cdot\hat{\mathbf{S}}_{2}+\left(\hat{S}_{1}^{z}-\hat{S}_{2}^{z}\right)\right],\label{delH_c}
\end{align}
where the coupling coefficients $\lambda_{\sigma,\sigma'}$ are defined
as follows
\begin{equation}
\lambda_{\sigma,\sigma'}=\frac{\sigma_{\epsilon}^{2}t^{2}}{2}\frac{d^{2}\Omega_{\sigma,\sigma'}^{-1}}{d\epsilon^{2}}|_{\epsilon=\bar{\epsilon}}.
\end{equation}
Thus, the effect of the charge noise may be viewed as providing a
perturbation $\delta\hat{H}_{\mathrm{c}}$ on the level of model Hamiltonian.
It turns out that the main effects are in the exchange interaction
and the local magnetic fields.

\section{Controlled phase gate\label{Sec_III}}

Among 
\begin{figure}
\begin{centering}
\includegraphics[width=8.6cm]{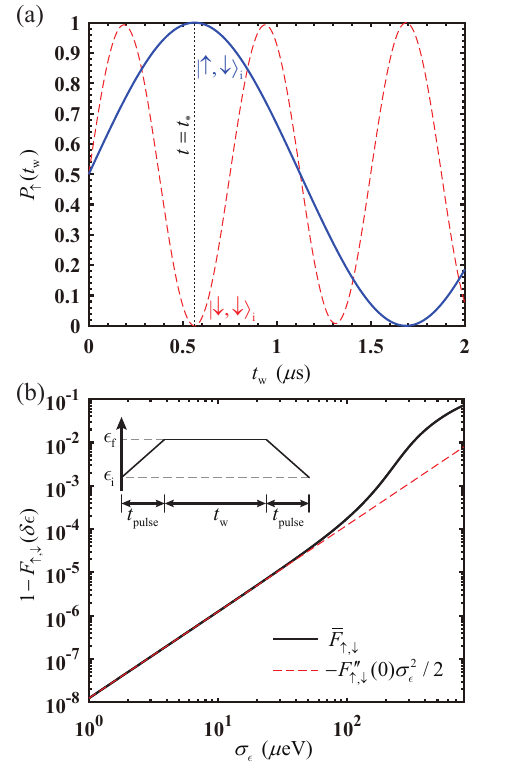}
\par\end{centering}
\caption{(a) Spin-up fraction as a function of waiting time for two different
initial states. The vertical dotted line depict $t_{\mathrm{w}}=t_{\mathrm{*}}$,
where an ideal controlled phase (CPHASE) gate could be realized. (b)
Gate infidelity integrated over the distribution $p(\epsilon)$ as
a function of the standard deviation $\sigma_{\epsilon}$. The black
solid and red dashed curves are the exact as well as approximated
results. Inset: schematic plot of the detuning as a function of time
during the implementation of the CPHASE gate. Parameters used in the
calculations: $t=4\mu\mathrm{eV}$, $\omega=150\mu\mathrm{eV}$, $\delta\omega=0.04\mu\mathrm{eV}$,
$U=220\mathrm{meV}$, $\epsilon_{\mathrm{f}}=218610\mu\mathrm{eV}$,
$\epsilon_{\mathrm{i}}=\epsilon_{\mathrm{f}}-1\mathrm{meV}$, and
$\bar{\epsilon}=\epsilon_{\mathrm{f}}$.\label{FIG_CPHASEgate}}
\end{figure}
 various two-qubit gates, the two most commonly considered ones are
the swap gate and the controlled phase (CPHASE) gate. Although other
types of two-qubit gates such as the controlled-not gate could also
be implemented with an additional driving on the conditional transition
\citep{2021_PRB_Heinz}, the exchange interaction has been exploited
for realizing the CPHASE gate \citep{2018_PRB_Russ}, thus the CPHASE
gate represents a good example to investigate the effect of charge
noise. Depending on the values of $J_{\mathrm{eff}}$ and $\delta\omega$,
the DQD system could realize both types of quantum gates \citep{2015_PRB_Veldhorst,2019_Nature_Huang,2018_Nature_Watson}.
In particular, for weak exchange interaction $J_{\mathrm{eff}}\ll\delta\omega$,
the eigenstates of the DQD in the spin anti-parallel subspace are
hardly affected by the tunneling coupling and resemble the product
states $|\sigma,\bar{\sigma}\rangle$ ($\bar{\sigma}$ denotes the
opposite direction to $\sigma$), thus the net effect of the exchange
interaction was to induce an additional frequency shifts for $|\sigma,\bar{\sigma}\rangle$
compared with the triplet states $|\sigma,\sigma\rangle$. Time-accumulated
dynamical phase due to this frequency shift then leads to a CPHASE
gate \citep{2011_PRB_Meunier}:
\begin{equation}
\mathbf{U}_{\mathrm{cp}}=\left[\begin{array}{cccc}
1 & 0 & 0 & 0\\
0 & e^{i\phi_{1}} & 0 & 0\\
0 & 0 & e^{i\phi_{2}} & 0\\
0 & 0 & 0 & 1
\end{array}\right],\label{Ucp}
\end{equation}
where the basis ordering for the above matrix is $\{|\uparrow,\uparrow\rangle,|\uparrow,\downarrow\rangle,|\downarrow,\uparrow\rangle,|\downarrow,\downarrow\rangle\}$.
For $\phi_{1}+\phi_{2}=\pi$, the CPHASE gate $\mathbf{U}_{\mathrm{cp}}$
is called a controlled $\pi$-phase gate. 

Let us consider the implementation of the CPHASE gate with the non-adiabatic
scheme \citep{2018_PRB_Russ}, consisting of non-adiabatically pulsing
the detuning from $\epsilon_{\mathrm{i}}$ to $\epsilon_{\mathrm{f}}$
and keeping the system evolving at $\epsilon_{\mathrm{f}}$ for a
period of waiting time $t_{\mathrm{w}}$ (cf. inset of FIG. \ref{FIG_CPHASEgate}).
The non-adiabatic pulsing of $\epsilon$ requires that the time $t_{\mathrm{pulse}}$,
during which $\epsilon$ changes from $\epsilon_{\mathrm{i}}$ to
$\epsilon_{\mathrm{f}}$, should satisfy $t_{\mathrm{pulse}}\ll J_{\mathrm{eff}}/\delta\omega^{2}$
\citep{2018_PRB_Russ}. For $J_{\mathrm{eff}}\sim3\mathrm{MHz}$ and
$\delta\omega\sim10\mathrm{MHz}$ this amounts to require $t_{\mathrm{pulse}}\ll30\mathrm{ns}$,
which was already demonstrated in experiments, for example charge
shuttling along a linear array of 9 quantum dots can be achieved in
$50\mathrm{ns}$ \citep{2019_NatCommun_Mills}.

For simplification, one may rewrite the effective Hamiltonian under
a rotating frame defined by the following time-dependent unitary transformation
\citep{2018_PRB_Russ}
\begin{equation}
\hat{R}(t)=e^{-i\omega(\hat{S}_{1}^{z}+\hat{S}_{2}^{z})t/2},
\end{equation}
Thus the transformed effective Hamiltonian reads
\begin{align}
\hat{H}_{\mathrm{rot}} & \equiv\hat{R}^{\dagger}(t)\hat{H}_{\mathrm{eff}}\hat{R}(t)-i\hat{R}^{\dagger}(t)\partial_{t}\hat{R}(t)\nonumber \\
 & =\epsilon_{\mathrm{ref}}-\frac{J_{\mathrm{eff}}}{4}+J_{\mathrm{eff}}\hat{\mathbf{S}}_{1}\cdot\hat{\mathbf{S}}_{2}+\frac{\delta\omega'}{2}\left(\hat{S}_{1}^{z}-\hat{S}_{2}^{z}\right),\label{Heff_rotframe}
\end{align}
Notice that Eq. (\ref{Heff_rotframe}) could be diagonalized with
a unitary transformation via $\hat{H}_{\mathrm{rot}}=\hat{U}_{\mathrm{rot}}(\epsilon)\hat{E}_{\mathrm{rot}}\hat{U}_{\mathrm{rot}}(\epsilon)^{\dagger}$,
where $\hat{E}_{\mathrm{rot}}$ is a diagonal matrix with diagonal
elements the energies in the ascending order. By introducing Pauli
matrices defined on the spin parallel (P) as well as spin anti-parallel
(AP) subspaces:
\begin{equation}
\hat{\tau}_{\mathrm{P}}^{0}=\sum_{\sigma\in\{\uparrow,\downarrow\}}|\sigma,\sigma\rangle\langle\sigma,\sigma|,
\end{equation}
\begin{equation}
\hat{\tau}_{\mathrm{AP}}^{x}=|\uparrow,\downarrow\rangle\langle\downarrow,\uparrow|+|\uparrow,\downarrow\rangle\langle\downarrow,\uparrow|,
\end{equation}
\begin{equation}
\hat{\tau}_{\mathrm{AP}}^{z}=|\uparrow,\downarrow\rangle\langle\uparrow,\downarrow|-|\downarrow,\uparrow\rangle\langle\downarrow,\uparrow|,
\end{equation}
then 
\begin{equation}
\hat{U}_{\mathrm{rot}}(\epsilon)=\hat{\tau}_{\mathrm{P}}^{0}+\cos\frac{\theta(\epsilon)}{2}\hat{\tau}_{\mathrm{AP}}^{x}-\sin\frac{\theta(\epsilon)}{2}\hat{\tau}_{\mathrm{AP}}^{z},
\end{equation}
where the mixing angle $\theta(\epsilon)=\arctan J_{\mathrm{eff}}/\delta\omega'$.
Notice that the unitary transformation $\hat{U}_{\mathrm{rot}}(\epsilon)$
defined above is also Hermitian. 

Since the system was initialized and readout at the detuning $\epsilon_{\mathrm{i}}$,
the natural basis for the discussion would be the eigenstates of $\hat{H}_{\mathrm{rot}}(\epsilon_{\mathrm{i}})$.
Notice that the tunneling coupling $t$ would have much less effect
on those eigenstates at the initial detuning $\epsilon_{\mathrm{i}}$
that is usually away from the $\mathrm{S}(0,2)$-$(1,1)$ anti-crossing
point, if $|t|^{2}/(U|\delta\omega|)\ll1$ \citep{2011_PRB_Meunier}.
In fact, one can show that to leading order of this small parameter,
$\sin\theta_{\mathrm{i}}\simeq4|t|/|\delta\omega|\cdot|t|/U$. Thus
the eigenstates in the spin anti-parallel subspace at the initial
detuning would be very close to the the product basis $|\sigma,\bar{\sigma}\rangle$,
e.g., $|\uparrow,\downarrow\rangle+2|t|^{2}/(|\delta\omega|U)|\downarrow,\uparrow\rangle\simeq|\uparrow,\downarrow\rangle$.
Therefore, we also denote the eigenbasis at $\epsilon_{\mathrm{i}}$
using the same notation $|\sigma,\sigma'\rangle_{\mathrm{i}}$ except
with a subscript $\mathrm{i}$ indicating $\epsilon_{\mathrm{i}}$.
By this notation we introduced the basis $|\uparrow,\uparrow\rangle_{\mathrm{i}}$,
$|\downarrow,\uparrow\rangle_{\mathrm{i}}$, $|\uparrow,\downarrow\rangle_{\mathrm{i}}$,
$|\downarrow,\downarrow\rangle_{\mathrm{i}}$, hereafter we refer
this basis as the $i$-basis.

The corresponding time-evolution operator when the system was biased
to $\epsilon_{\mathrm{f}}$ is written as follows\begin{widetext}
\begin{align}
\hat{U}(\epsilon_{\mathrm{f}},t_{\mathrm{w}})= & e^{-i\hat{H}_{\mathrm{rot}}(\epsilon_{\mathrm{f}})t_{\mathrm{w}}}\nonumber \\
= & \hat{U}_{\mathrm{rot}}(\epsilon_{\mathrm{f}})e^{-i\hat{E}_{\mathrm{rot}}(\epsilon_{\mathrm{f}})t_{\mathrm{w}}}\hat{U}_{\mathrm{rot}}(\epsilon_{\mathrm{f}})^{\dagger}\nonumber \\
= & \left[\begin{array}{cccc}
1 & 0 & 0 & 0\\
0 & \cos^{2}\frac{\theta_{\mathrm{i}}-\theta_{\mathrm{f}}}{2}e^{-i\varphi_{-}}+\sin^{2}\frac{\theta_{\mathrm{i}}-\theta_{\mathrm{f}}}{2}e^{-i\varphi_{+}} & \frac{\sin(\theta_{\mathrm{i}}-\theta_{\mathrm{f}})}{2}\left(e^{-i\varphi_{-}}-e^{-i\varphi_{+}}\right) & 0\\
0 & \frac{\sin(\theta_{\mathrm{i}}-\theta_{\mathrm{f}})}{2}\left(e^{-i\varphi_{-}}-e^{-i\varphi_{+}}\right) & \sin^{2}\frac{\theta_{\mathrm{i}}-\theta_{\mathrm{f}}}{2}e^{-i\varphi_{-}}+\cos^{2}\frac{\theta_{\mathrm{i}}-\theta_{\mathrm{f}}}{2}e^{-i\varphi_{+}} & 0\\
0 & 0 & 0 & 1
\end{array}\right].\label{Umat}
\end{align}
\end{widetext}where in the last line the matrix was written under
the $i$-basis defined above, the mixing angle $\theta_{\mathrm{i}/\mathrm{f}}=\theta(\epsilon_{\mathrm{i}/\mathrm{f}})$
and the time-accumulated dynamical phases are
\begin{equation}
\varphi_{\pm}=\frac{1}{2}\left[\pm\sqrt{\delta\omega'(\epsilon_{\mathrm{f}}){}^{2}+J_{\mathrm{eff}}(\epsilon_{\mathrm{f}})^{2}}-J_{\mathrm{eff}}(\epsilon_{\mathrm{f}})\right]t_{\mathrm{w}}.
\end{equation}

In order to realize the ideal CPHASE gate, one would require the waiting
time $t_{\mathrm{w}}$ to satisfy $\varphi_{+}=\varphi_{-}+2n\pi$
with $n$ an integer. For $n=1$ this yields the following condition
\begin{equation}
t_{\mathrm{w}}=t_{*}\equiv\frac{2\pi}{\sqrt{\delta\omega'(\epsilon_{\mathrm{f}}){}^{2}+J_{\mathrm{eff}}(\epsilon_{\mathrm{f}})^{2}}}.
\end{equation}
When $t_{\mathrm{w}}=t_{*}$, the evolution matrix Eq. (\ref{Umat})
becomes diagonal under the $i$-basis,
\begin{equation}
\hat{U}(\epsilon_{\mathrm{f}})=\hat{\tau}_{\mathrm{P},i}^{0}-e^{i\pi\sin\theta_{\mathrm{f}}}\hat{\tau}_{\mathrm{AP},i}^{0},\label{UcpEff}
\end{equation}
where $\hat{\tau}_{\mathrm{A},i}^{0}$ and $\hat{\tau}_{\mathrm{AP},i}^{0}$
in Eq. (\ref{UcpEff}) denote the identity Pauli matrix defined with
respect to the $i$-basis. Notice that the mixing angle $\theta_{\mathrm{f}}$
depends on the detuning $\epsilon_{\mathrm{f}}$, thus if $\epsilon_{\mathrm{f}}$
is chosen to be $\epsilon_{*}$ such that $\theta_{\mathrm{f}}=\pi/6$,
then Eq. (\ref{UcpEff}) realizes the inverse of $\hat{U}_{\mathrm{cp}}$
defined in Eq. (\ref{Ucp}) with $\phi_{1}=\phi_{2}=\pi/2$. We verified
the implementation of the CPHASE gate by calculating the spin-up fraction
of the second quantum dot \citep{2015_Nature_Veldhorst}
\begin{equation}
P_{\uparrow}(t_{\mathrm{w}})=\sum_{s\in\{\uparrow,\downarrow\}}|_{\mathrm{i}}\langle s,\uparrow|\tilde{R}_{y}^{(2)}(\frac{\pi}{2})\hat{U}(\epsilon_{*},t_{\mathrm{w}})\tilde{R}_{x}^{(2)}(\frac{\pi}{2})|\psi_{0}\rangle|^{2},
\end{equation}
where $\hat{R}_{\alpha}^{(j)}(\theta)=\exp[-i\hat{S}_{j}^{\alpha}\theta]$
is the rotation operator for the $j$-th quantum dot with rotation
angle $\theta$ and rotation axis along the $\alpha$-direction, and
$\tilde{O}=\hat{U}_{\mathrm{rot}}(\epsilon_{\mathrm{i}})\hat{O}\hat{U}_{\mathrm{rot}}^{\dagger}(\epsilon_{\mathrm{i}})$
thus $\tilde{R}_{\alpha}^{(j)}(\theta)$ is the corresponding rotation
operator with spin orientation defined according to the $i$-basis.
The results are shown in FIG. \ref{FIG_CPHASEgate}(a). In particular,
when the waiting time $t_{\mathrm{w}}$ is chosen as $t_{*}$, the
time evolution following initial state $|\uparrow,\downarrow\rangle_{\mathrm{i}}$
and $|\downarrow,\downarrow\rangle_{\mathrm{i}}$ leads to opposite
behavior in $P_{\uparrow}$, consistent with previous studies \citep{2015_Nature_Veldhorst}.

The noise on $\epsilon$ bias the detuning from the ideal value $\epsilon_{\mathrm{f}}^{\mathrm{ideal}}=\epsilon_{*}$
such that $\theta_{\mathrm{f}}=\pi/6$ to $\epsilon_{\mathrm{f}}^{\mathrm{actual}}=\epsilon_{*}+\delta\epsilon$,
which leads to errors in implementing the CPHASE gate. To quantify
this error we calculated the Ulhmann fidelity between the state $\hat{U}(\epsilon_{*})|\uparrow,\downarrow\rangle_{\mathrm{i}}$
and $\hat{U}(\epsilon_{*}+\delta\epsilon)|\uparrow,\downarrow\rangle_{\mathrm{i}}$,
which can be regarded as a gate fidelity \citep{2012_PRA_Magesan},
the result is given by
\begin{equation}
F_{\uparrow,\downarrow}(\delta\epsilon)=\cos^{2}\frac{\delta\varphi}{2}+\cos^{2}\delta\theta\sin^{2}\frac{\delta\varphi}{2},
\end{equation}
where 
\begin{equation}
\delta\varphi=\varphi_{+}(\epsilon_{*}+\delta\epsilon)-\varphi_{-}(\epsilon_{*}+\delta\epsilon),
\end{equation}
\begin{equation}
\delta\theta=\theta(\epsilon_{*})-\theta(\epsilon_{*}+\delta\epsilon).
\end{equation}
The corresponding gate infidelity $1-F_{\uparrow,\downarrow}$ averaged
over the distribution $p(\epsilon)$ with $\bar{\epsilon}=\epsilon_{*}$
is shown in FIG. \ref{FIG_CPHASEgate}(b), for small value of $\sigma_{\epsilon}$
the infidelity can be well described by the second order derivative
of $F_{\uparrow,\downarrow}$, i.e., $F_{\uparrow,\downarrow}(0)+F_{\uparrow,\downarrow}^{''}(0)\sigma_{\epsilon}^{2}/2$.
The $\sigma_{\epsilon}^{2}$ scaling at weak noise, i.e., small $\sigma_{\epsilon}$,
is a result of the Gaussian distribution that assumed for $p(\epsilon)$.
While the deviation from $\sigma_{\epsilon}^{2}$ scaling for a stronger
noise is due to the failure of second order perturbative expansion
on $F_{\uparrow,\downarrow}$.

\section{Randomized benchmarking\label{Sec_IV}}

Instead 
\begin{figure}
\begin{centering}
\includegraphics[width=8.6cm]{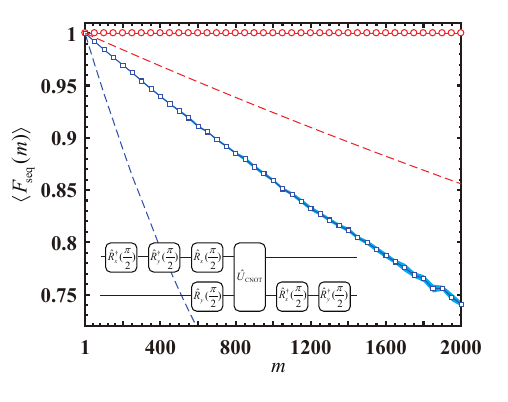}
\par\end{centering}
\caption{Two-qubit randomized benchmarking of the controlled phase gate (CPHASE)
subject to an error in detuning $\delta\epsilon_{\mathrm{f}}=\epsilon_{\mathrm{f}}^{\mathrm{actual}}-\epsilon_{\mathrm{f}}^{\mathrm{ideal}}$.
Red circles (blue squares) show the averaged sequential fidelity following
the standard (interleaved) RB. The blue shaded region shows the standard
deviation for the interleaved RB simulation. The dashed curves show
the corresponding RB simulation including a gate-independent dephasing
effect. Inset: Decomposition of CPHASE gates in terms of Clifford
gates, the gates on the left act first while the first (second) line
represents the control (target) qubit. Parameters used in calculation:
$t=4\mu\mathrm{eV}$, $\epsilon_{\mathrm{ref}}=0$, $\omega=150\mu\mathrm{eV}$,
$\delta\omega=0.04\mu\mathrm{eV}$, $U=220\mathrm{meV}$, $1-p_{\mathrm{dp}}=10^{-4}$,
$\delta\epsilon=28\mu\mathrm{eV}$, and $K=500$.\label{FIG_IRB}}
\end{figure}
 of the gate fidelity $F_{\uparrow,\downarrow}$, a more robust and
systematic way of characterizing the noise 
\begin{figure}
\begin{centering}
\includegraphics[width=8.6cm]{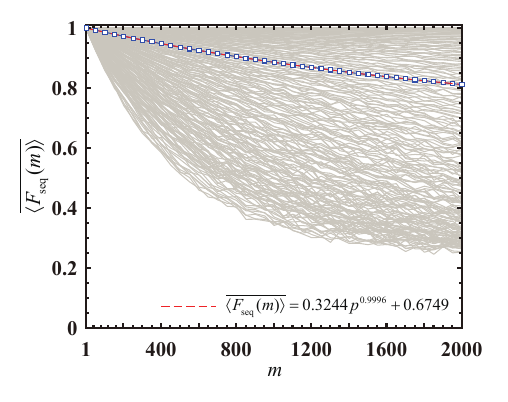}
\par\end{centering}
\caption{The sequential fidelity averaged over the charge noise distribution
$p(\epsilon)$. The blue squares are the numerical data point while
the red dashed curve is the fit to the zero-th order RB model. The
gray curves represents 200 RB simulations with different detuning
errors $\delta\epsilon$ sampled uniformly within $\epsilon_{*}\pm3\sigma_{\epsilon}$.
Parameters used in the calculations: $t=4\mu\mathrm{eV}$, $\epsilon_{\mathrm{ref}}=0$,
$\omega=150\mu\mathrm{eV}$, $\delta\omega=0.04\mu\mathrm{eV}$, $U=220\mathrm{meV}$,
$\sigma_{\epsilon}=28\mu\mathrm{eV}$, and $K=500$.\label{FIG_RBAVG}}

\end{figure}
effect on the CPHASE gate is through the interleaved randomized benchmarking
(IRB) \citep{2005_JPhysB_Emerson,2012_PRA_Magesan,2012_PRL_Magesan,2014_NJP_Wallman,2014_PRA_Epstein}.
Usually, in a randomized benchmarking (RB) test on an $n$-qubits
system, a sequence of $m$ quantum gates are sampled randomly from
the $n$-qubits Clifford group $\mathcal{C}_{n}$. The system is then
evolved by first subjecting to the $m$ gates then applied with a
recovery gate, which in the ideal noiseless case would return the
system to its initial state. Thus the wave function overlapping between
the ideal and actually evolved final state provides a measure on the
fidelity loss due to the noise. 

The wave function overlapping is formally defined through the sequential
fidelity $F_{\mathrm{seq}}(m)$, given by
\begin{equation}
F_{\mathrm{seq}}(m)=\mathrm{Tr}\left\{ \hat{E}_{\xi}\left[\prod_{i=1}^{m+1}\Lambda_{i}\circ\mathcal{C}_{i}\right]\hat{\rho}_{0}\right\} ,
\end{equation}
Here, $\{\hat{E}_{\xi}\}$ is a set of measurement operators with
possible outcomes $\{\xi\}$, the complete positive and trace-preserving
(CPTP) quantum operation $\Lambda_{i}$ describes the noisy effects
including decoherence during the implementation of Clifford gate $\mathcal{C}_{i}$,
while $\hat{\rho}_{0}$ is the initial state of the benchmarked quantum
system.

Here we consider the randomized benchmarking (RB) based on the Clifford
gates, the ideal CPHASE gate Eq. (\ref{Ucp}) with $\phi_{1}=\phi_{2}=\pi/2$
can be decomposed in terms of two-qubit Clifford gates as shown in
FIG. \ref{FIG_IRB} in the sense of having identical Pauli transfer
matrices. In FIG. \ref{FIG_IRB} we show the (SRB) standard as well
as (IRB) interleaved RB simulation results under a fixed detuning
error $\delta\epsilon=28\mu\mathrm{eV}$. The IRB simulation assuming
an additional gate-independent dephasing process with $\Lambda[\hat{\rho}]=p_{\mathrm{dp}}\hat{\rho}+2(1-p_{\mathrm{dp}})\sum_{i=1,2}\tilde{S}_{i}^{z}\hat{\rho}\tilde{S}_{i}^{z}$
and $1-p_{\mathrm{dp}}=10^{-4}$ was also shown. Fitting the IRB data
to the zero-th order fitting model \citep{2012_PRL_Magesan}
\begin{equation}
\langle F_{\mathrm{seq}}(m)\rangle=Ap^{m}+B,\label{Fitmodel}
\end{equation}
gives $p=0.9998$, where the bracket $\langle F_{\mathrm{seq}}(m)\rangle$
means to average $F_{\mathrm{seq}}(m)$ over the random sequences
of gates with length $m$. Thus the gate error associated with the
CPHASE gate was extracted to be $1.5\times10^{-4}$, which is slightly
larger than the $10^{-5}$ infidelity estimated from the spin-up fraction
calculation.

The effect of charge noise on the RB simulation can be obtained by
further averaging $\langle F_{\mathrm{seq}}(m)\rangle$ over the distribution
$p(\epsilon)$ with $\bar{\epsilon}=\epsilon_{*}$, i.e.,
\begin{equation}
\overline{\langle F_{\mathrm{seq}}(m)\rangle}=\int d\epsilon p(\epsilon)\langle F_{\mathrm{seq}}(m)\rangle,
\end{equation}
The resulting decaying behavior of $\overline{\langle F_{\mathrm{seq}}(m)\rangle}$
is shown in FIG. \ref{FIG_RBAVG} for $\sigma_{\epsilon}=28\mu\mathrm{eV}$.
It turns out that the zero-th order fitting model Eq. (\ref{Fitmodel})
still provides quantitative well fit to the simulated data for the
realistic parameters used in this study. However, we notice that in
general $\overline{\langle F_{\mathrm{seq}}(m)\rangle}$ may deviates
from the exponential decay behavior predicted by Eq. (\ref{Fitmodel}),
e.g., power-law decay of the sequential fidelity had been reported
in previous studies for single qubit due to the noise in qubit frequency
\citep{2015_PRA_Fogarty}.

In the RB simulation we have focused on the the reported realistic
parameters and did not considered the full parameter space. Clearly,
as suggested from the gate fidelity calculation in Sec. \ref{Sec_III},
as long as $\sigma_{\epsilon}$ becomes larger enough one may expect
the deviation from the approximation $F_{\uparrow,\downarrow}(0)+F_{\uparrow,\downarrow}^{''}(0)\sigma_{\epsilon}^{2}/2$.
Also, in the simulation here we have neglected the effect of decoherence
and focused on the influence of control errors related to the detuning,
thus the sequential fidelity decays much slower than those found in
experiments. As can be seen in FIG. \ref{FIG_IRB}, inclusion of pure
dephasing on the qubit could leads to fast decay of the sequential
fidelity, which is qualitatively consistent with recent experiment
on isotopically purified silicon system, where the sequential fidelity
on a controlled rotation gate already decays for a sequence with $60$
gates \citep{2019_Nature_Huang}.

\section{Conclusion}

In this work, we have explored the effect of charge noise on the controlled
phase gate in a double quantum dot. By assuming quasistatic noise
on the dot detuning, we have derived simple expressions for the Hamiltonian,
the exchange interaction, as well as the gate fidelity that are valid
under weak noise relevant for experiments. Those analyses were also
supplied with the randomized benchmarking simulations, showing that
an exponential decay of the sequential fidelity could be still valid
under noise average. Those results would be helpful for a better understanding
of charge noise effect and the design of high fidelity two-qubit gates
in semiconductor based spin qubits.
\begin{acknowledgments}
The author thanks Stefano Chesi for very helpful discussions. Y.F.
acknowledges support from NSFC (Grant No. 12005011) and Yunnan Fundamental
Research Projects (Grant No. 202201AU070118).
\end{acknowledgments}

\bibliographystyle{apsrev4-1}
\bibliography{References}

\end{document}